\documentclass[aps, pra, reprint]{revtex4-1}

\usepackage[bookmarks=false]{hyperref}
\usepackage{graphicx}
\usepackage[utf8]{inputenc}
\usepackage{xcolor}
\usepackage{amsmath, amssymb}

\begin{document}

\title{Berezinskii–Kosterlitz–Thouless Transition in Two-Dimensional Dipolar Stripes}

\author{Raúl Bombín}
\email{raul.bombin@upc.edu}
\affiliation{Departament de F\'{i}sica, Universitat Polit\`{e}cnica de Catalunya, Campus Nord B4-B5, E-08034, Barcelona, Spain}

\author{Ferran Mazzanti}
\email{ferran.mazzanti@upc.edu}
\affiliation{Departament de F\'{i}sica, Universitat Polit\`{e}cnica de Catalunya, Campus Nord B4-B5, E-08034, Barcelona, Spain}

\author{Jordi Boronat}
\email{jordi.boronat@upc.edu}
\affiliation{Departament de F\'{i}sica, Universitat Polit\`{e}cnica de Catalunya, Campus Nord B4-B5, E-08034, Barcelona, Spain}

\begin{abstract}
A two-dimensional quantum system of dipoles, with a polarization angle not 
perpendicular to the plane, shows a transition from a gas to a stripe phase.
We have studied the thermal properties of these two phases using the path 
integral Monte Carlo (PIMC) method. By simulating the thermal density matrix, 
PIMC provides exact results for magnitudes of interest such as the 
superfluid fraction and the one-body density matrix. As it is well known, in 
two dimensions the superfluid-to-normal phase transition follows the 
Berezinskii–Kosterlitz–Thouless (BKT) scenario. Our results show that both the 
anisotropic gas and the stripe phases follow the BKT scaling laws. At fixed 
density and increasing the tilting angle, the transition temperature decreases 
in going from the gas to the stripe phase. Superfluidity in the perpendicular 
direction to the stripes is rather small close to the critical temperature but 
it becomes larger at lower temperatures, mainly close to the transition to the 
gas. Our results are in qualitative agreement with the supersolidity observed 
recently in a quasi-one-dimensional array of dipolar droplets.
\end{abstract}

\maketitle
\section{Introduction}
The achievement of supersolidity as a new state of matter has been a 
long-standing topic since it was  theoretically predicted in the sixties of the 
past century~\cite{Andreev1969}. A supersolid state is produced when 
two  U(1) symmetries are simultaneously broken: the first 
one related to  the presence of spatial long-range order and the 
second one to 
the emergence of a global phase giving rise to a superfluid state. 
The most natural candidate to be a supersolid is solid $^4$He, due to its 
extreme quantum character. However, and after a big excitement produced some 
years ago, the most accurate data available to the date seem to exclude this 
possibility~\cite{Kim2012}. 

The difficulties in finding a stable condensed-matter supersolid state has 
moved its research to metastable systems which can exhibit the same properties.
In  recent years, the most fruitful tool to this end has been 
the versatile setup of ultracold quantum gases in the quantum  
degenerate regime. Although conventional dilute Bose Einstein Condensate gases 
(BEC's) do not break translational symmetry and thus are not good 
candidates for supersolid phases, some progress has been recently achieved by 
taking advantage of more exotic interactions. The first evidence of 
supersolidity came in 2017 almost simultaneously from two different experiments 
in reduced geometries. In the first one, a spin-orbit coupled system was shown 
to break 
translational symmetry in a two-dimensional configuration~\cite{Li2013} 
whereas, in the second one, this effect was achieved by  
coupling a Bose–Einstein condensate to the modes of two optical 
cavities~\cite{Leonard2017}. 
Still in the context of ultracold gases, dipolar systems have been postulated 
as good candidates to the supersolid state. In fact, supersolid signatures have 
been observed by several 
groups~\cite{Tanzi2019,Chomaz2019,Bottcher2019}, following previous theoretical 
work~\cite{Roccuzzo2019}. Recently, the gapless Goldstone excitation  has also been 
measured for the same system~\cite{Tanzi2019b,Guo2019,natale2019}.

\begin{center}
\begin{figure}[tb]
\includegraphics[width=0.9\linewidth]{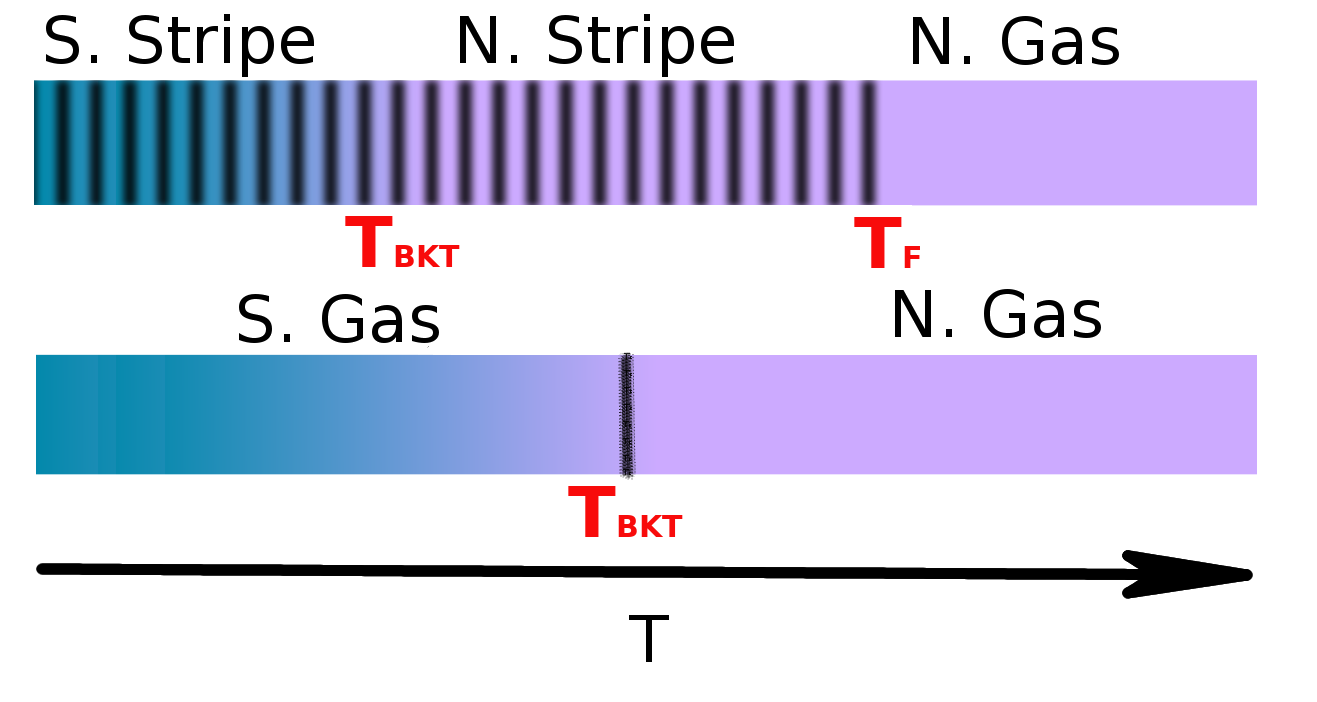}
\caption{Phase transitions in dipolar stripes compared with the dipolar gas. N 
and S labels stand for normal and superfluid phases, respectively.}
\label{fig:cartoon_phases}
\end{figure}
\end{center}

In a previous work~\cite{Bombin2017}, the superfluid properties of the different 
phases  of a dipolar system in two dimensions were studied at zero 
temperature. The stripe phase, that appears for certain densities and 
tilting angles, was shown to exhibit the characteristics which define a supersolid state. 
Recently, similar results have been reported for the equivalent system in the 
lattice \cite{Bandyopadhyay2019}.
At finite temperature, there is not condensate but quasi-condensate 
reflected in an algebraic decay of the one-body density matrix.  
The transition between the superfluid (with quasi-off-diagonal long range order) 
and normal phases follows the   Berezinskii, Kosterlitz and Thouless 
(BKT) theory~\cite{Berezinskii1971,Kosterlitz1973}. This transition 
has been studied in many different systems such as Helium 
films~\cite{Agnolet1989,Ceperley1989,Gordillo1998}, 
Coulomb layers~\cite{Minnhagen1987}, and 
ultracold gases in pancake geometries~\cite{Desbuquois2012,Ota2018,Murthy2015}. 
It has also been shown that the BKT scenario stands even when disorder is introduced in the 
system~\cite{Carleo2013,Maccari2018}.

In this paper, we study the superfluid-to-normal phase transition in a system 
of two-dimensional bosonic dipoles performing first principles 
Path Integral Monte Carlo (PIMC) simulations. The particular case in which all 
the dipoles are polarized along the direction perpendicular to the plane, which 
constitutes the isotropic case, was already studied by Filinov \textit{et 
al.}~\cite{Filinov2010}. Here, we focus on the more general case in which 
dipoles are 
polarized in an arbitrary direction, within the stability limit,   
and show that the BKT scaling stands despite of the anisotropy induced by 
the dipolar interaction. We determine the critical temperature $T_{\rm{BKT}}$ 
in both the gas and stripe phases. As schematically illustrated in 
Fig.~\ref{fig:cartoon_phases}, $T_{\rm{BKT}}$  for the stripe phase is smaller 
than  $T_{\rm{BKT}}$ for the gas, at the same density. Increasing 
further the 
temperature, we observe that the normal stripes melt towards an anisotropic 
gas. 

\section{Method}

The system under study is composed of  $N$ identical dipolar
bosons of mass $m$ moving on the $XY$ plane. An external field (electric or 
magnetic)  in the XZ plane polarizes all the dipoles along the same direction in 
space, forming an angle $\alpha$ with respect to the Z axis.  The model
Hamiltonian describing the system reads
\begin{equation}
H = -\frac{\hbar^2}{2m}\sum_{j=1}^N \nabla_j^2
+ \frac{C_{dd}}{4\pi} \sum_{i<j}^N
\left[ \frac{1 - 3\lambda^2 \cos^2\theta_{ij}}{r_{ij}^3} \right] \ ,
\label{Hamiltonian}
\end{equation}
with $\lambda=\sin\alpha$,
and $(r_{ij},\theta_{ij})$ the polar coordinates of $\mathbf{r}_{ij}$.
The strength of the dipolar interaction is encoded in the constant $C_{dd}$  
and is proportional to the square of the (electric or
magnetic) dipole moment of each particle.  Similarly to 
previous works, we employ dipolar units~\cite{Astrakharchik2007,Bombin2017}, with the 
characteristic dipolar length $r_0=m C_{dd}/(4\pi\hbar^2)$ 
and dipolar energy  $\varepsilon_0 = \hbar^2/(mr_0^2)$ 
that allows for writing the Hamiltonian in dimensionless units. 
In the following, temperatures will be expressed also in units of 
$\varepsilon_0$. The system is 
stable towards collapse as long as the tilting angle $\alpha$ is smaller than 
the critical value $\alpha_c \simeq 0.61$. Our simulations are carried out in 
a rectangular box, with periodic boundary conditions (PBC), to 
correctly commensurate the stripes~\cite{Bombin2017}, similarly to what
is made in the simulation of crystals. 

For a given Hamiltonian, the PIMC method provides exact results (within some 
statistical noise) for the energy, structure and superfluidity of a Bose fluid 
or solid. It has been widely used in the past to study the BKT transition, for 
instance  in two-dimensional liquid $^4$He~\cite{Ceperley1989,Ceperley1995} and 
in dipoles with dipolar moments perpendicular to the 
plane~\cite{Filinov2010}. Going down in temperature, and mainly close to the 
critical temperature, the PIMC simulation requires of a good action to reduce 
the number of imaginary-time steps (\textit{beads}) representing each atom 
(\textit{polymer}) to a manageable level. To this end,  we use the fourth order 
Chin's action~\cite{Takahashi1984,Takahashi1984a,Chin2004,Chin2002}, that can 
be made to work effectively up to sixth order for the energy estimation by 
optimizing its control  
parameters~\cite{Sakkos2009}. Efficiency in the sampling of permutations is 
also fundamental to obtain accurate results for the one-body density matrix 
and superfluid densities. To get it right we use the worm algorithm, that has 
proven its accuracy in different systems~\cite{Boninsegni2006}.

At odds with what happens in three-dimensional systems, the superfluid fraction 
performs an abrupt universal jump~\cite{Nelson1977} at the critical temperature 
$T_c$. Near $T_c$,  the BKT theory predicts that the correlation length has 
an essential singularity  $\xi(T)\sim e^{a/t^{1/2}}$, with $t = 
(T/T_c -1)$ and $a$ being a non-universal parameter depending on density 
and on the microscopic properties of the particular system under study~\cite {Kosterlitz1974}. 
Due to the use of a finite number of particles $N$, within a finite-size box with PBC, we do not have direct access to the critical temperature in the thermodynamic limit ($T_c(\infty)$) 
but rather to an estimation $T_c(L)$, with $L=\sqrt{N/n}$. As usual in finite-size scaling 
analysis of simulations close to the critical point, one identifies $T_c(L)$ 
with the temperature that makes $\xi(T_c(L))=L$. Therefore, the scaling law of 
the critical temperature with the size of the box can be written as~\cite{Filinov2010} 
\begin{equation}
T_c(L) = T_c(\infty) + \frac{b}{\ln^2(L\sqrt{n})} \ ,
\label{eqn.Tcscaling}
\end{equation}
with $b$ a non-universal constant. On the other hand, the jump that the 
superfluid density performs at the critical temperature $T_c$ follows the universal 
relation~\cite{Nelson1977} 
\begin{equation}
\frac{n_s(T_c,L)}{n} = \frac{2mk_B}{\pi\hbar^2} \frac{T_c}{n} \ ,
\label{eqn.UniversalJUMP}
\end{equation}
with $k_B$ the Boltzmann constant. 

\section{Results}
\subsection{Superfluid fraction}
In order to determine the critical temperature at which the superfluid-to-normal phase transition occurs, we need to evaluate the superfluid density. In the PIMC method, this is done through the well known \textit{winding number} estimator~\cite{Pollock1987},
\begin{equation}
\frac{n_s}{n} = \frac{m k_B T}{N \hbar^2}\langle \textbf{W}^2\rangle \ ,
\label{eqn:Winding}
\end{equation}
where \textbf{W} is the  winding number. 
\subsubsection{BKT scaling of the gas phase}

\begin{center}
\begin{figure}[tb]
\includegraphics[width=0.99\linewidth]{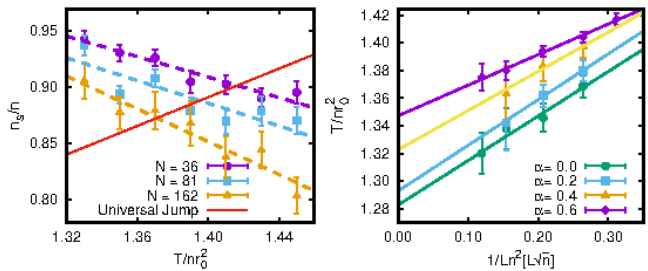}
\caption{Left panel: superfluid fraction as a function of temperature for 
different system sizes at density $nr_0^2 = 25$ and tilt angle $\alpha = 0.6$.  
Points are MC results, dashed lines are linear fits to PIMC data and the solid 
line 
is the universal jump of Eq. (\ref{eqn.UniversalJUMP}). The crossing points 
between the  lines and the universal jump give the critical 
temperatures $T_C(L)$. Right panel: scaling of the critical Temperature 
$T_c(L)$ 
with the system size, as given by Eq. (\ref{eqn.Tcscaling}), at the same 
density and for different polarization angles. Points are PIMC data and solid 
lines are linear fits.}
\label{fig:gas_scaling}
\end{figure}
\end{center}

Using the superfluid densities, 
calculated with the estimator (\ref{eqn:Winding}) at different 
temperatures and system sizes, and taking advantage of the universal relations 
of equations \eqref{eqn.Tcscaling} and \eqref{eqn.UniversalJUMP}, one can 
obtain the superfluid-to-normal critical temperature. 
We start studying the transition in the gas phase at different densities and 
tilting angles. In Fig. \ref{fig:gas_scaling}, we show our 
PIMC results for the superfluid fraction $n_s/n$ at 
a density $nr_0^2 = 25$. In the left panel of this figure, we 
show our results for 
a tilting angle $\alpha = 0.6$, close to the border 
of stability of the gas at zero temperature~\cite{Macia2014}. 
The critical temperature for a given system size $T_c(L)$ is 
determined as the crossing point between the universal BKT jump of 
Eq.~(\ref{eqn.UniversalJUMP}) and the superfluid density for that system 
size. On the right panel of the same figure, we show how the  scaling   
\eqref{eqn.Tcscaling} is used  to obtain the critical temperature in 
the thermodynamic limit. The analysis for different values of the tilting angle 
$\alpha = 0$, $0.2$, $0.4$, and $0.6$ reveals that the BKT scaling stands when  
anisotropy is present in the system. 

Our results for $\alpha=0$, 
corresponding to the isotropic gas, reproduce the PIMC estimations obtained by 
Filinov \textit{et al.}~\cite{Filinov2010}.
In that work, it was found a non-monotonic behavior 
of the critical temperature as a function of the density~\cite{Filinov2010}. 
The critical temperature, in units of density $T_c/nr_0^2$,  increases  
at low densities and, above a characteristic value ($ 1 < nr_0^2 < 4$), the 
behavior is the opposite. Filinov \textit{et al.}~\cite{Filinov2010} attribute 
this change to the appearance of the roton in the 
quasi-particle spectrum, which is observed to emerge around 
$nr_0^2=1$~\cite{Macia2009,Filinov2010,Astrakharchik2007}.
We have studied how the tilting angle ($\alpha>0$) influences this behavior 
by calculating $T_c$ at  low ($nr_0^2 = 0.01$) and high ($nr_0^2=25$) 
densities, as shown in   Table~\ref{table:1}.  The 
behavior of  $T_c/nr_0^2$   with the tilting 
angle is the opposite for densities $0.01$ and $25$: 
increasing $\alpha$ reduces (increases) the critical temperature at low 
(high) density.  In both cases, though,  the growth of $\alpha$ translates 
into an effective reduction of the interaction strength  since the 
$s$-wave scattering length for a given tilting angle is well approximated 
by~\cite{Macia2011},    
\begin{equation}
a_s(\lambda) \simeq e^{2\gamma}\left(1-\frac{3\lambda^2}{2}\right) \ ,
\label{gammafun}
\end{equation}
with gamma the Euler's Gamma constant. According to Eq. (\ref{gammafun}), the 
scattering length for dipolar interaction decreases when $\alpha$ 
increases. In agreement with the isotropic case~\cite{Filinov2010}, the 
effective reduction of the interaction strength lowers $T_c$ at low densities, 
where the excitation spectrum is phononic, but increases it at high densities, 
when rotons dominate. 

\begin{table}[tb]
\centering
\begin{tabular}{c c c c || c c c c } 
\hline
\multicolumn{8}{c}{Gas Phase}\\
\hline
$nr_0^2$& $\alpha$ & $T_c/nr_0^2$ $[\varepsilon_0]$ & $n_s/n(T_c)$& $nr_0^2$& 
$\alpha$ & 
$T_c/nr_0^2$ $[\varepsilon_0]$ &$n_s/n(T_c)$\\ 
 \hline
 0.01 & 0.0 & 1.316(6) & 0.838(4) & 25 & 0.0 & 1.282(8) & 0.816(6)\\ 
 0.01 & 0.2 & 1.317(3) & 0.838(6) & 25 & 0.2 & 1.292(5) & 0.823(4) \\
 0.01 & 0.4 & 1.29(11) & 0.821(6) & 25 & 0.4 & 1.322(1) & 0.842(3) \\
 0.01 & 0.6 & 1.263(13)& 0.804(8)  & 25 & 0.6 & 1.347(3) & 0.858(2) \\
 128 & 0.4 & 1.04(4) & 0.66(3) & 256 & 0.4 & 0.82(3) & 0.52(2) \\ 
 \hline
 \\
\hline 
 \multicolumn{8}{c}{Stripe Phase}\\
\hline
$nr_0^2$& $\alpha$ & $T_c/nr_0^2$ $[\varepsilon_0]$ & $n_s/n(T_c)$& $nr_0^2$& 
$\alpha$ & 
 $T_c/nr_0^2$ $[\varepsilon_0]$  &$n_s/n(T_c)$\\ 
 \hline
 128 & 0.6 & 0.60(7) & 0.38(4) & 256 & 0.6 & 0.49(4) & 0.31(3) \\
 \hline
\end{tabular}
\caption{BKT critical temperatures (in dipolar units) for different values of 
the 
density $nr_0^2$ and 
tilting angle $\alpha$, in both the gas and stripe phases. The 
superfluid fraction at the critical temperature is 
evaluated through Eq.~\eqref{eqn.UniversalJUMP}. Figures in parenthesis are the 
estimated errors. }
\label{table:1}
\end{table}

\subsubsection{BKT scaling of the stripe phase}

\begin{center}
\begin{figure}[tb]
\includegraphics[width=0.99\linewidth]{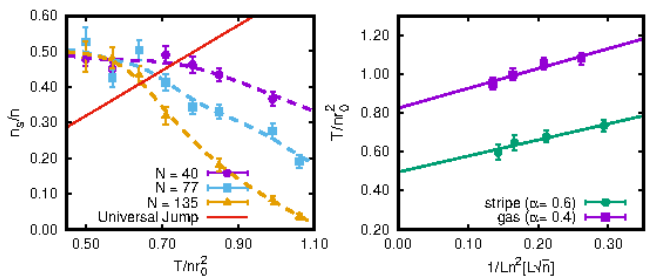}
\caption{Left panel: superfluid fraction as a function of the temperature for 
different system 
sizes, at density $nr_0^2 = 256$ and tilting angle $\alpha = 0.6$, 
corresponding 
to 
the stripe phase.  Points are PIMC results, dashed lines are guides to the eye, 
and the solid 
line is the universal jump (\ref{eqn.UniversalJUMP}). Right panel: scaling of 
the critical temperature $T_c(L)$ with the system 
size, as given by Eq. (\ref{eqn.Tcscaling}), at the same density and for two 
tilting 
angles: $\alpha = 0.4$ (gas) and $0.6$ (stripe). Points are PIMC data 
and solid lines are linear fits.}
\label{fig:stripe_scaling}
\end{figure}
\end{center}

The stripe phase is of particular relevance in our study since it has been 
reported to be superfluid in the zero-temperature limit~\cite{Bombin2017}. The 
simultaneous existence of spatial long-range order (in all but one direction of 
the space) and off-diagonal long-range order makes this phase to be close to 
the pursued supersolid state of matter. A relevant issue in this 
discussion is whether the BKT scaling, that we have shown to hold for the 
anisotropic gas, stands also for the stripe phase. In 
Fig.~\ref{fig:stripe_scaling}, we show PIMC results for the superfluid 
fraction at a density 
$nr_0^2= 256$ and tilting angle $\alpha = 0.6$ where the stripe phase is 
stable~\cite{Bombin2017}. In the left panel, 
we show the behavior of the superfluid fraction as a function of temperature 
and for different number of particles in the simulation box. 
As in the gas phase, the crossing of this lines with 
the universal jump law of Eq. (\ref{eqn.UniversalJUMP}) allows us to extract 
the 
critical temperature for a given system size $T_c(L)$. In the right panel, we 
compare the scaling of these critical temperatures for the stripe phase with 
the ones 
obtained for the same density but  at a  smaller tilting angle
$\alpha =0.4$ where the gas phase is the stable one.
As one can see, the  BKT scaling holds in both cases,
and thus one can apply it to estimate the 
critical temperature in the thermodynamic limit. 

One could think  
that the stripe phase is composed 
of quasi-one-dimensional channels, which dominate the superfluid signal, 
in such a way that 
 the superfluidity in stripes follow the one-dimensional 
scaling law instead of the BKT one. In the next section we show that this is not 
the case, and thus only the BKT scenario is plausible with our results (see section \ref{sec:1D}).

For temperatures lower than $T_c$, the superfluid fraction shows  a plateau 
around a value which is in agreement with the zero-temperature result derived 
previously using the diffusion Monte Carlo method~\cite{Bombin2017},  
$\left[\frac{n_s}{n}\right]_{\alpha=0.6}^{nr_0^2=256}=0.54(5)$.

In Table~\ref{table:1}, we report the results for the critical temperature and 
superfluid fraction at $T_c$ of the stripe phase with $\alpha=0.6$ and 
densities $nr_0^2=128$ and $256$. By increasing the density, the critical 
temperature in the stripe phase decreases in a similar form to what
has been previously obtained for the gas at high density. However, if the tilting 
angle increases, at fixed density, and crosses from the gas to the stripe phase 
both the superfluid fraction and the critical temperature decrease (see for 
instance data at $nr_0^2=128$ in Table~\ref{table:1}). In other 
words, superfluidity in stripes is thermally more fragile than in the gas 
phase.  The winding number estimator for superfluidity (\ref{eqn:Winding}) can 
be split into the $X$ and $Y$ directions corresponding to the stripe  
orientation and its perpendicular one, respectively. At $T_c$, the superfluid 
fraction in the $Y$ direction for a finite $N$ value is $<5$\% and decreases 
with $T$ faster than 
the one along the stripe direction. As it was 
observed previously~\cite{Bombin2017}, the superfluidity across the stripes 
depends strongly on the tilting angle, keeping the density fixed, reaching 
values $\sim 100$\% close to the gas-stripe phase transition line but 
decreasing fast when entering the deep stripe region. 

\subsubsection{Non Luttinger Liquid behaviour of the stripe phase.}
\label{sec:1D}

One may wonder if the stripe phase at finite temperature might be considered as 
an ensemble of one-dimensional systems. If this were the case, our data should 
accommodate to the predictions of the Luttinger Liquid (LL) theory. Although  
one-dimensional systems do not show superfluidity in the thermodynamic 
limit, one can still see a non-zero superfluid fraction in a finite system of 
length $L$. For a one-dimensional liquid, described by Luttinger theory, the 
superfluid fraction for a Galilean invariant system is predicted to scale with 
the system size 
as~\cite{VranjesMarkic2018}
\begin{equation}
\frac{n_s}{n}=\frac{\gamma}{4}\frac{\left.|\Theta_3^{\prime\prime}(0,e^{
-\gamma/2})\right|}{\Theta_3(0,e^{-\gamma/2})}
\label{eqn:sup_1D}
\end{equation}
where $\Theta_3(z,q)$ is the Theta function, $\Theta_3^{\prime\prime}(z,q) = 
d^2\Theta_3(z,q)/dz^2$, and $\gamma = \frac{m k_B T L}{\hbar^2 n_l}$ with $n_l$ 
the linear density.  

\begin{figure}[tb]
\includegraphics[width=0.75\linewidth]{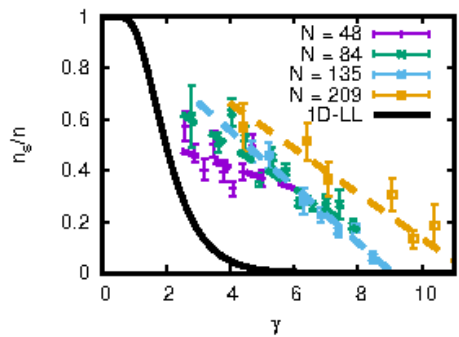}
\caption{
Superfluid fraction of the stripe phase for different number of particles 
as a function of the scaling parameter of Luttinger theory $\gamma$ for density $nr_0^2=128$ and tilting angle $\alpha=0.6$. Solid black line corresponds to the Luttinger liquid prediction of Eq.~\eqref{eqn:sup_1D}.
As it can be seen, there is not collapse of the data to a single line. }
\label{fig:1D}
\end{figure}

In Fig.~\ref{fig:1D}, we show that the data for the stripe phase ($nr_0^2 = 128$ 
and $\alpha = 0.6$) do not collapse to a single line when doing the scaling 
with $\gamma$, with a lineal density $n_l=14.6(3)$ obtained from $n_l=N/(L 
N_s)$ with $N_s$ the number of stripes in the simulation box containing $N$ 
particles. In the same figure we show the prediction of the Luttinger Liquid theory (black line), whose comparison with our results hints that the superfluid signal in the stripes is more robust against system size and temperature (encoded in the parameter $\gamma$) than what the Luttinger theory predicts for a 1D system.
Therefore, we conclude that the stripe phase of a two-dimensional dipolar 
system cannot be 
considered as an ensemble of one-dimensional Luttinger liquids. This result 
is in agreement with the analysis of simulation data 
of the one-body density matrix of the stripe phase at zero 
temperature~\cite{Bombin2017}.  

\subsection{The One-Body density matrix}

To get a deeper insight in the supersolid properties of the stripe phase, we 
have calculated the one-body density matrix (OBDM),
\begin{equation}
n_1(\mathbf{r}^\prime_1,\mathbf{r}_1) = \frac{V}{Z}\int d 
\mathbf{r}_2 \ldots \mathbf{r}_N \, \rho(\mathbf{R}^\prime,\mathbf{R}),
\label{eqn:OBDM}
\end{equation}
with  $\mathbf{R}=\{\mathbf{r}_1,\mathbf{r}_2, \ldots,\mathbf{r}_N\}$, 
$\mathbf{R}^\prime=\{\mathbf{r}^\prime_1,\mathbf{r}_2,\ldots, 
\mathbf{r}_N\}$, $\rho(\mathbf{R}^\prime,\mathbf{R})$ the thermal density 
matrix, and $Z$ the partition function. As 
it is well known, in 2D systems there is a condensate fraction only in the 
$T=0$ limit. This condensate fraction, which means that the system has 
off-diagonal long-range order, is obtained from the asymptotic constant value 
of $n_1(\mathbf{r}^\prime_1,\mathbf{r}_1)$ at large distances. For $T \leq 
T_c$,  
$n_1(\mathbf{r}^\prime_1,\mathbf{r}_1)$ decays with a power law instead, 
pointing to what is generally termed as quasi-condensate. In contrast, for 
$T>T_c$ 
the decay turns out to be exponential, as it corresponds to a normal phase.

\begin{center}
\begin{figure}[b]
\includegraphics[width=0.95\linewidth]{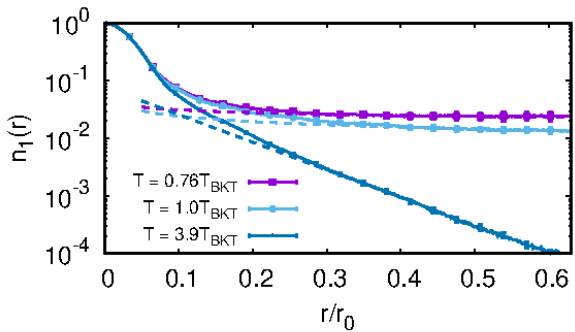}
\caption{One-body density matrix of the stripe phase  ($nr_0^2 = 
128$ and  $\alpha = 0.6$) at 
different temperatures, above and below the transition temperature 
$T_c$. The straight lines correspond to the asymptotic behavior when 
$r \to \infty$. }
\label{fig:OBDM}
\end{figure}
\end{center}

In Fig.~\ref{fig:OBDM}, we show PIMC results for the 
OBDM in the stripe phase ($nr_0^2 = 128$, $\alpha 
= 0.6$) at different temperatures. Below the BKT 
transition temperature, 
the long-range behavior of the OBDM is well captured with a fit of 
the form $n_1(r)\sim 
r^{-\eta}$. The value of the exponent $\eta$ is given by the BKT theory,  
\begin{equation}
\eta = (m k_B T)/(2 \pi \hbar n_s),
\label{eqn:BKT_exp}
\end{equation}
becoming maximal at the critical 
point, $\eta_c = 1/4$. As we can see in Fig.~\ref{fig:OBDM}, the algebraic 
decay of the PIMC results below $T_c$  reproduce the BKT 
prediction. When the stripes become normal, the OBDM changes dramatically and 
we 
clearly see an exponential decay.

\subsection{Stripe melting}

\begin{figure}[t]
\includegraphics[width=0.95\linewidth]{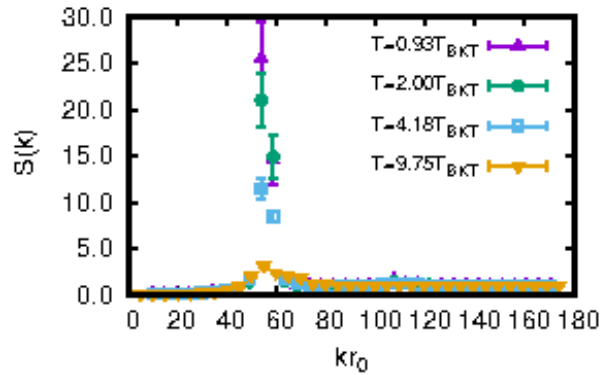}
\caption{
Evolution with the temperature of the static structure factor $S_y(k)$ in the 
stripe phase at $nr_0^2=128$ and $\alpha=0.6$.}
\label{fig:sk}
\end{figure}

When temperature is increased beyond $T_c$, the stripe phase still persists as 
the ground state of the system, but being a normal phase (non-superfluid). 
Under these conditions, the static structure factor still shows a clear Bragg peak in 
the transverse direction ($Y$) pointing to the stability of the 
stripes~\cite{Macia2012}. Thus this is an interesting quantity if one wants to estimate, 
the critical temperature at which the stripe phase melts towards the gas one. To study this, we evaluate the the static structure factor for wave vectors perpendicular ($Y$) to 
the stripe direction ($X$),
\begin{equation}
S_y(k) = \frac{1}{N Z} \langle \hat{\rho}_{-\mathbf{k}_y}  
\hat{\rho}_{\mathbf{k}_y} \rangle \ ,
\end{equation}
with $\hat{\rho}_{\mathbf{k}_y}=\sum_{i=1}^{N} e^{i \mathbf{k}_y\cdot 
\mathbf{r}_i}$ the density-fluctuation operator. In Fig.\ref{fig:sk}, we show 
results of $S_y(k)$, for a characteristic point of the phase diagram 
where the system is in the stripe phase, as a 
function of the temperature. 

The Bragg peak that appears at a characteristic $\mathbf{k}_y$ signals the 
periodic pattern of the stripes in their transverse direction. This large peak, 
which increases with the number of particles 
$N$~\cite{Macia2009,Macia2012}, is the best signature 
of the stripe order. When the temperature increases, the strength of the peak 
decreases due to the increase of the thermal motion.  At the largest temperature 
reported in Fig.\ref{fig:sk}, the Bragg peak has disappeared pointing to its 
melting to a gas. Notice that no equivalent peak appears at any $T$ in the $X$ 
direction.

However, the 
localization decreases progressively with $T$ until we observe their melting at 
a temperature $T/\simeq 10\, T_{\textsc{BKT}}$

\begin{figure}[t]
\includegraphics[width=0.9\linewidth]{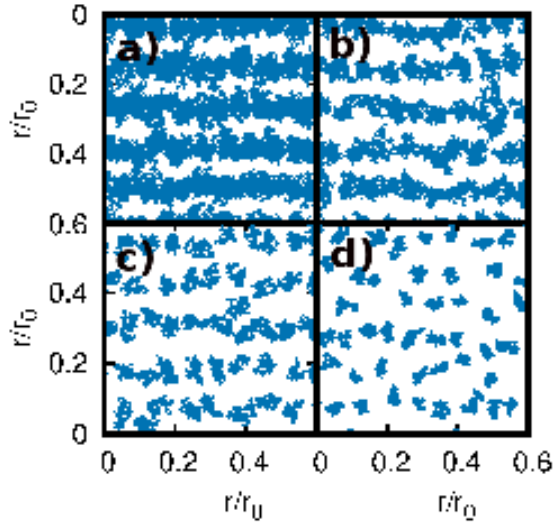}
\caption{
Snapshots of the PIMC simulations of the stripe phase for increasing 
temperatures at $nr_0^2=128$ and $\alpha=0.6$. The temperature $T$ 
increases from a) to d) panels. The values of $T$ are the same than 
in Fig.~\ref{fig:sk}.}
\label{fig:snaps}
\end{figure}

The evolution of the stripe structure can also be qualitatively analyzed by 
looking at the spatial distribution of particles in the PIMC simulation. In 
Fig.~\ref{fig:snaps}, 
we show snapshots to show this evolution with increasing $T$. In the PIMC 
framework, each particle is represented by a polymer with an averaged size 
proportional to its 
quantum delocalization. At temperatures below $T_c$, one can see from the 
snapshots that there are paths connecting the different linear structures 
(stripes); when these crossing paths are of the length of the simulation box 
there is a nonzero winding number in that direction and the superfluid fraction 
is finite. In the second frame of Fig.~\ref{fig:snaps}, this transverse paths 
have nearly disappeared and also in the $X$ direction the interconnections are 
not very abundant. In the third frame, we still observe the characteristic 
order 
of stripes but dislocations between the 
different lines starts to be apparent. This effect has been deeply studied in 
Refs.~\cite{wu2016liquid,mendoza2017quantum} and now our microscopic 
simulations confirm these predictions. Finally, the last frame corresponds to a 
temperature where the stripe structure is no more present because it has melted 
to a (normal) gas.

\section{Conclusions}

In conclusion, we have carried out a complete study of the BKT transition in  
anisotropic 2D systems of quantum dipoles. Using the BKT theory we have 
estimated the superfluid-to-normal phase transition critical temperature 
at different densities and tilting angles. At fixed density, and increasing 
the tilting angle, we observe the transition from a gas to a stripe phase 
with a decrease on the critical 
temperature in the stripe case. In spite of this reduction, which makes the 
supersolid phase of stripes less stable against thermal fluctuations  
than the gas, the superfluid 
signal is clear below $T_c$. The long-range behavior of the OBDM is also 
consistent with the BKT prediction. Interestingly, our PIMC results on the 
superfluid fraction shows that its value in the transverse direction is still 
finite but small ($<5$\%) close to $T_c$ and that at lower temperatures, and 
mainly close to the transition line to the gas, its value is much larger, 
almost $100$\%. This result is qualitatively similar to recent experiments 
in which a dipolar droplet system, arranged in a quasi-one-dimensional array, has 
shown superfluid signatures across the drops~\cite{Tanzi2019,Chomaz2019,Bottcher2019}. 
Regarding two-dimensional dipolar systems, similar predictions about the existence
of a superfluid stripe phase have been recently reported for the equivalent
system in the lattice \cite{Bandyopadhyay2019}. Therefore, the quantum dipolar phases seem now the best suited candidates for the realization of the pursued supersolid state of matter. Finally, it is also worth mentioning that a superfluid stripe phase has
been studied in the Hubbard model with an isotropic long-range interaction. In this
case, the rotational symmetry is broken spontaneously by the interplay between the
long-range character of the inter-particle interaction considered with the lattice, that
forces the atoms to occupy certain lattice positions in order to minimize the energy \cite{Masella2019}.

\acknowledgments{
This work has been supported by the Ministerio de
Economia, Industria y Competitividad (MINECO, Spain) under grant No. 
FIS2017-84114-C2-1-P.
 }


\bibliography{refs}

\begin{thebibliography}{44}%
\makeatletter
\providecommand \@ifxundefined [1]{%
 \@ifx{#1\undefined}
}%
\providecommand \@ifnum [1]{%
 \ifnum #1\expandafter \@firstoftwo
 \else \expandafter \@secondoftwo
 \fi
}%
\providecommand \@ifx [1]{%
 \ifx #1\expandafter \@firstoftwo
 \else \expandafter \@secondoftwo
 \fi
}%
\providecommand \natexlab [1]{#1}%
\providecommand \enquote  [1]{``#1''}%
\providecommand \bibnamefont  [1]{#1}%
\providecommand \bibfnamefont [1]{#1}%
\providecommand \citenamefont [1]{#1}%
\providecommand \href@noop [0]{\@secondoftwo}%
\providecommand \href [0]{\begingroup \@sanitize@url \@href}%
\providecommand \@href[1]{\@@startlink{#1}\@@href}%
\providecommand \@@href[1]{\endgroup#1\@@endlink}%
\providecommand \@sanitize@url [0]{\catcode `\\12\catcode `\$12\catcode
  `\&12\catcode `\#12\catcode `\^12\catcode `\_12\catcode `\%12\relax}%
\providecommand \@@startlink[1]{}%
\providecommand \@@endlink[0]{}%
\providecommand \url  [0]{\begingroup\@sanitize@url \@url }%
\providecommand \@url [1]{\endgroup\@href {#1}{\urlprefix }}%
\providecommand \urlprefix  [0]{URL }%
\providecommand \Eprint [0]{\href }%
\providecommand \doibase [0]{http://dx.doi.org/}%
\providecommand \selectlanguage [0]{\@gobble}%
\providecommand \bibinfo  [0]{\@secondoftwo}%
\providecommand \bibfield  [0]{\@secondoftwo}%
\providecommand \translation [1]{[#1]}%
\providecommand \BibitemOpen [0]{}%
\providecommand \bibitemStop [0]{}%
\providecommand \bibitemNoStop [0]{.\EOS\space}%
\providecommand \EOS [0]{\spacefactor3000\relax}%
\providecommand \BibitemShut  [1]{\csname bibitem#1\endcsname}%
\let\auto@bib@innerbib\@empty
\bibitem [{\citenamefont {Andreev}\ and\ \citenamefont
  {Lifshitz}(1971)}]{Andreev1969}%
  \BibitemOpen
  \bibfield  {author} {\bibinfo {author} {\bibfnamefont {A.~F.}\ \bibnamefont
  {Andreev}}\ and\ \bibinfo {author} {\bibfnamefont {I.~M.}\ \bibnamefont
  {Lifshitz}},\ }\href {\doibase 10.1070/pu1971v013n05abeh004235} {\bibfield
  {journal} {\bibinfo  {journal} {Soviet Physics Uspekhi}\ }\textbf {\bibinfo
  {volume} {13}},\ \bibinfo {pages} {670} (\bibinfo {year} {1971})}\BibitemShut
  {NoStop}%
\bibitem [{\citenamefont {Kim}\ and\ \citenamefont {Chan}(2012)}]{Kim2012}%
  \BibitemOpen
  \bibfield  {author} {\bibinfo {author} {\bibfnamefont {D.~Y.}\ \bibnamefont
  {Kim}}\ and\ \bibinfo {author} {\bibfnamefont {M.~H.}\ \bibnamefont {Chan}},\
  }\href {\doibase 10.1103/PhysRevLett.109.155301} {\bibfield  {journal}
  {\bibinfo  {journal} {Physical Review Letters}\ }\textbf {\bibinfo {volume}
  {109}},\ \bibinfo {pages} {155301} (\bibinfo {year} {2012})}\BibitemShut
  {NoStop}%
\bibitem [{\citenamefont {Li}\ \emph {et~al.}(2013)\citenamefont {Li},
  \citenamefont {Martone}, \citenamefont {Pitaevskii},\ and\ \citenamefont
  {Stringari}}]{Li2013}%
  \BibitemOpen
  \bibfield  {author} {\bibinfo {author} {\bibfnamefont {Y.}~\bibnamefont
  {Li}}, \bibinfo {author} {\bibfnamefont {G.~I.}\ \bibnamefont {Martone}},
  \bibinfo {author} {\bibfnamefont {L.~P.}\ \bibnamefont {Pitaevskii}}, \ and\
  \bibinfo {author} {\bibfnamefont {S.}~\bibnamefont {Stringari}},\ }\href@noop
  {} {\bibfield  {journal} {\bibinfo  {journal} {Physical Review Letters}\
  }\textbf {\bibinfo {volume} {110}},\ \bibinfo {pages} {235302} (\bibinfo
  {year} {2013})}\BibitemShut {NoStop}%
\bibitem [{\citenamefont {L{\'{e}}onard}\ \emph {et~al.}(2017)\citenamefont
  {L{\'{e}}onard}, \citenamefont {Morales}, \citenamefont {Zupancic},
  \citenamefont {Esslinger},\ and\ \citenamefont {Donner}}]{Leonard2017}%
  \BibitemOpen
  \bibfield  {author} {\bibinfo {author} {\bibfnamefont {J.}~\bibnamefont
  {L{\'{e}}onard}}, \bibinfo {author} {\bibfnamefont {A.}~\bibnamefont
  {Morales}}, \bibinfo {author} {\bibfnamefont {P.}~\bibnamefont {Zupancic}},
  \bibinfo {author} {\bibfnamefont {T.}~\bibnamefont {Esslinger}}, \ and\
  \bibinfo {author} {\bibfnamefont {T.}~\bibnamefont {Donner}},\ }\href
  {\doibase 10.1038/nature21067} {\bibfield  {journal} {\bibinfo  {journal}
  {Nature}\ }\textbf {\bibinfo {volume} {543}},\ \bibinfo {pages} {87}
  (\bibinfo {year} {2017})}\BibitemShut {NoStop}%
\bibitem [{\citenamefont {Tanzi}\ \emph
  {et~al.}(2019{\natexlab{a}})\citenamefont {Tanzi}, \citenamefont {Lucioni},
  \citenamefont {Fam{\`{a}}}, \citenamefont {Catani}, \citenamefont {Fioretti},
  \citenamefont {Gabbanini}, \citenamefont {Bisset}, \citenamefont {Santos},\
  and\ \citenamefont {Modugno}}]{Tanzi2019}%
  \BibitemOpen
  \bibfield  {author} {\bibinfo {author} {\bibfnamefont {L.}~\bibnamefont
  {Tanzi}}, \bibinfo {author} {\bibfnamefont {E.}~\bibnamefont {Lucioni}},
  \bibinfo {author} {\bibfnamefont {F.}~\bibnamefont {Fam{\`{a}}}}, \bibinfo
  {author} {\bibfnamefont {J.}~\bibnamefont {Catani}}, \bibinfo {author}
  {\bibfnamefont {A.}~\bibnamefont {Fioretti}}, \bibinfo {author}
  {\bibfnamefont {C.}~\bibnamefont {Gabbanini}}, \bibinfo {author}
  {\bibfnamefont {R.~N.}\ \bibnamefont {Bisset}}, \bibinfo {author}
  {\bibfnamefont {L.}~\bibnamefont {Santos}}, \ and\ \bibinfo {author}
  {\bibfnamefont {G.}~\bibnamefont {Modugno}},\ }\href {\doibase
  10.1103/PhysRevLett.122.130405} {\bibfield  {journal} {\bibinfo  {journal}
  {Physical Review Letters}\ }\textbf {\bibinfo {volume} {122}},\ \bibinfo
  {pages} {130405} (\bibinfo {year} {2019}{\natexlab{a}})}\BibitemShut
  {NoStop}%
\bibitem [{\citenamefont {Chomaz}\ \emph {et~al.}(2019)\citenamefont {Chomaz},
  \citenamefont {Petter}, \citenamefont {Ilzh{\"{o}}fer}, \citenamefont
  {Natale}, \citenamefont {Trautmann}, \citenamefont {Politi}, \citenamefont
  {Durastante}, \citenamefont {van Bijnen}, \citenamefont {Patscheider},
  \citenamefont {Sohmen}, \citenamefont {Mark},\ and\ \citenamefont
  {Ferlaino}}]{Chomaz2019}%
  \BibitemOpen
  \bibfield  {author} {\bibinfo {author} {\bibfnamefont {L.}~\bibnamefont
  {Chomaz}}, \bibinfo {author} {\bibfnamefont {D.}~\bibnamefont {Petter}},
  \bibinfo {author} {\bibfnamefont {P.}~\bibnamefont {Ilzh{\"{o}}fer}},
  \bibinfo {author} {\bibfnamefont {G.}~\bibnamefont {Natale}}, \bibinfo
  {author} {\bibfnamefont {A.}~\bibnamefont {Trautmann}}, \bibinfo {author}
  {\bibfnamefont {C.}~\bibnamefont {Politi}}, \bibinfo {author} {\bibfnamefont
  {G.}~\bibnamefont {Durastante}}, \bibinfo {author} {\bibfnamefont {R.~M.~W.}\
  \bibnamefont {van Bijnen}}, \bibinfo {author} {\bibfnamefont
  {A.}~\bibnamefont {Patscheider}}, \bibinfo {author} {\bibfnamefont
  {M.}~\bibnamefont {Sohmen}}, \bibinfo {author} {\bibfnamefont {M.~J.}\
  \bibnamefont {Mark}}, \ and\ \bibinfo {author} {\bibfnamefont
  {F.}~\bibnamefont {Ferlaino}},\ }\href {\doibase 10.1103/PhysRevX.9.021012}
  {\bibfield  {journal} {\bibinfo  {journal} {Physical Review X}\ }\textbf
  {\bibinfo {volume} {9}},\ \bibinfo {pages} {021012} (\bibinfo {year}
  {2019})}\BibitemShut {NoStop}%
\bibitem [{\citenamefont {B{\"{o}}ttcher}\ \emph {et~al.}(2019)\citenamefont
  {B{\"{o}}ttcher}, \citenamefont {Schmidt}, \citenamefont {Wenzel},
  \citenamefont {Hertkorn}, \citenamefont {Guo}, \citenamefont {Langen},\ and\
  \citenamefont {Pfau}}]{Bottcher2019}%
  \BibitemOpen
  \bibfield  {author} {\bibinfo {author} {\bibfnamefont {F.}~\bibnamefont
  {B{\"{o}}ttcher}}, \bibinfo {author} {\bibfnamefont {J.-N.}\ \bibnamefont
  {Schmidt}}, \bibinfo {author} {\bibfnamefont {M.}~\bibnamefont {Wenzel}},
  \bibinfo {author} {\bibfnamefont {J.}~\bibnamefont {Hertkorn}}, \bibinfo
  {author} {\bibfnamefont {M.}~\bibnamefont {Guo}}, \bibinfo {author}
  {\bibfnamefont {T.}~\bibnamefont {Langen}}, \ and\ \bibinfo {author}
  {\bibfnamefont {T.}~\bibnamefont {Pfau}},\ }\href {\doibase
  10.1103/PhysRevX.9.011051} {\bibfield  {journal} {\bibinfo  {journal}
  {Physical Review X}\ }\textbf {\bibinfo {volume} {9}},\ \bibinfo {pages}
  {011051} (\bibinfo {year} {2019})}\BibitemShut {NoStop}%
\bibitem [{\citenamefont {Roccuzzo}\ and\ \citenamefont
  {Ancilotto}(2019)}]{Roccuzzo2019}%
  \BibitemOpen
  \bibfield  {author} {\bibinfo {author} {\bibfnamefont {S.~M.}\ \bibnamefont
  {Roccuzzo}}\ and\ \bibinfo {author} {\bibfnamefont {F.}~\bibnamefont
  {Ancilotto}},\ }\href {\doibase 10.1103/PhysRevA.99.041601} {\bibfield
  {journal} {\bibinfo  {journal} {Physical Review A}\ }\textbf {\bibinfo
  {volume} {99}},\ \bibinfo {pages} {041601} (\bibinfo {year}
  {2019})}\BibitemShut {NoStop}%
\bibitem [{\citenamefont {Tanzi}\ \emph
  {et~al.}(2019{\natexlab{b}})\citenamefont {Tanzi}, \citenamefont {Roccuzzo},
  \citenamefont {Lucioni}, \citenamefont {Fam{\`{a}}}, \citenamefont
  {Fioretti}, \citenamefont {Gabbanini}, \citenamefont {Modugno}, \citenamefont
  {Recati},\ and\ \citenamefont {Stringari}}]{Tanzi2019b}%
  \BibitemOpen
  \bibfield  {author} {\bibinfo {author} {\bibfnamefont {L.}~\bibnamefont
  {Tanzi}}, \bibinfo {author} {\bibfnamefont {S.~M.}\ \bibnamefont {Roccuzzo}},
  \bibinfo {author} {\bibfnamefont {E.}~\bibnamefont {Lucioni}}, \bibinfo
  {author} {\bibfnamefont {F.}~\bibnamefont {Fam{\`{a}}}}, \bibinfo {author}
  {\bibfnamefont {A.}~\bibnamefont {Fioretti}}, \bibinfo {author}
  {\bibfnamefont {C.}~\bibnamefont {Gabbanini}}, \bibinfo {author}
  {\bibfnamefont {G.}~\bibnamefont {Modugno}}, \bibinfo {author} {\bibfnamefont
  {A.}~\bibnamefont {Recati}}, \ and\ \bibinfo {author} {\bibfnamefont
  {S.}~\bibnamefont {Stringari}},\ }\href {\doibase 10.1038/s41586-019-1568-6}
  {\bibfield  {journal} {\bibinfo  {journal} {Nature (accepted)}\ } (\bibinfo
  {year} {2019}{\natexlab{b}}),\ 10.1038/s41586-019-1568-6}\BibitemShut
  {NoStop}%
\bibitem [{\citenamefont {Guo}\ \emph {et~al.}(2019)\citenamefont {Guo},
  \citenamefont {B{\"o}ttcher}, \citenamefont {Hertkorn}, \citenamefont
  {Schmidt}, \citenamefont {Wenzel}, \citenamefont {B{\"u}chler}, \citenamefont
  {Langen},\ and\ \citenamefont {Pfau}}]{Guo2019}%
  \BibitemOpen
  \bibfield  {author} {\bibinfo {author} {\bibfnamefont {M.}~\bibnamefont
  {Guo}}, \bibinfo {author} {\bibfnamefont {F.}~\bibnamefont {B{\"o}ttcher}},
  \bibinfo {author} {\bibfnamefont {J.}~\bibnamefont {Hertkorn}}, \bibinfo
  {author} {\bibfnamefont {J.-N.}\ \bibnamefont {Schmidt}}, \bibinfo {author}
  {\bibfnamefont {M.}~\bibnamefont {Wenzel}}, \bibinfo {author} {\bibfnamefont
  {H.~P.}\ \bibnamefont {B{\"u}chler}}, \bibinfo {author} {\bibfnamefont
  {T.}~\bibnamefont {Langen}}, \ and\ \bibinfo {author} {\bibfnamefont
  {T.}~\bibnamefont {Pfau}},\ }\href {\doibase 10.1038/s41586-019-1569-5}
  {\bibfield  {journal} {\bibinfo  {journal} {Nature}\ }\textbf {\bibinfo
  {volume} {574}},\ \bibinfo {pages} {386} (\bibinfo {year}
  {2019})}\BibitemShut {NoStop}%
\bibitem [{\citenamefont {Natale}\ \emph {et~al.}(2019)\citenamefont {Natale},
  \citenamefont {van Bijnen}, \citenamefont {Patscheider}, \citenamefont
  {Petter}, \citenamefont {Mark}, \citenamefont {Chomaz},\ and\ \citenamefont
  {Ferlaino}}]{natale2019}%
  \BibitemOpen
  \bibfield  {author} {\bibinfo {author} {\bibfnamefont {G.}~\bibnamefont
  {Natale}}, \bibinfo {author} {\bibfnamefont {R.~M.~W.}\ \bibnamefont {van
  Bijnen}}, \bibinfo {author} {\bibfnamefont {A.}~\bibnamefont {Patscheider}},
  \bibinfo {author} {\bibfnamefont {D.}~\bibnamefont {Petter}}, \bibinfo
  {author} {\bibfnamefont {M.~J.}\ \bibnamefont {Mark}}, \bibinfo {author}
  {\bibfnamefont {L.}~\bibnamefont {Chomaz}}, \ and\ \bibinfo {author}
  {\bibfnamefont {F.}~\bibnamefont {Ferlaino}},\ }\href {\doibase
  10.1103/PhysRevLett.123.050402} {\bibfield  {journal} {\bibinfo  {journal}
  {Phys. Rev. Lett.}\ }\textbf {\bibinfo {volume} {123}},\ \bibinfo {pages}
  {050402} (\bibinfo {year} {2019})}\BibitemShut {NoStop}%
\bibitem [{\citenamefont {Bombin}\ \emph {et~al.}(2017)\citenamefont {Bombin},
  \citenamefont {Boronat},\ and\ \citenamefont {Mazzanti}}]{Bombin2017}%
  \BibitemOpen
  \bibfield  {author} {\bibinfo {author} {\bibfnamefont {R.}~\bibnamefont
  {Bombin}}, \bibinfo {author} {\bibfnamefont {J.}~\bibnamefont {Boronat}}, \
  and\ \bibinfo {author} {\bibfnamefont {F.}~\bibnamefont {Mazzanti}},\ }\href
  {\doibase 10.1103/PhysRevLett.119.250402} {\bibfield  {journal} {\bibinfo
  {journal} {Physical Review Letters}\ }\textbf {\bibinfo {volume} {119}},\
  \bibinfo {pages} {250402} (\bibinfo {year} {2017})}\BibitemShut {NoStop}%
\bibitem [{\citenamefont {Bandyopadhyay}\ \emph {et~al.}(2019)\citenamefont
  {Bandyopadhyay}, \citenamefont {Bai}, \citenamefont {Pal}, \citenamefont
  {Suthar}, \citenamefont {Nath},\ and\ \citenamefont
  {Angom}}]{Bandyopadhyay2019}%
  \BibitemOpen
  \bibfield  {author} {\bibinfo {author} {\bibfnamefont {S.}~\bibnamefont
  {Bandyopadhyay}}, \bibinfo {author} {\bibfnamefont {R.}~\bibnamefont {Bai}},
  \bibinfo {author} {\bibfnamefont {S.}~\bibnamefont {Pal}}, \bibinfo {author}
  {\bibfnamefont {K.}~\bibnamefont {Suthar}}, \bibinfo {author} {\bibfnamefont
  {R.}~\bibnamefont {Nath}}, \ and\ \bibinfo {author} {\bibfnamefont
  {D.}~\bibnamefont {Angom}},\ }\href {http://arxiv.org/abs/1906.07483}
  {\bibfield  {journal} {\bibinfo  {journal} {arXiv preprint arXiv:1906.07483}\
  } (\bibinfo {year} {2019})}\BibitemShut {NoStop}%
\bibitem [{\citenamefont {Berezinskii}(1971)}]{Berezinskii1971}%
  \BibitemOpen
  \bibfield  {author} {\bibinfo {author} {\bibfnamefont {V.~L.}\ \bibnamefont
  {Berezinskii}},\ }\href {\doibase 10.1051/jp3:1993206} {\bibfield  {journal}
  {\bibinfo  {journal} {Sov. Phys. JETP}\ }\textbf {\bibinfo {volume} {32}},\
  \bibinfo {pages} {493} (\bibinfo {year} {1971})},\ \Eprint
  {http://arxiv.org/abs/0512356} {0512356 [cond-mat]} \BibitemShut {NoStop}%
\bibitem [{\citenamefont {Kosterlitz}\ and\ \citenamefont
  {Thouless}(1973)}]{Kosterlitz1973}%
  \BibitemOpen
  \bibfield  {author} {\bibinfo {author} {\bibfnamefont {J.~M.}\ \bibnamefont
  {Kosterlitz}}\ and\ \bibinfo {author} {\bibfnamefont {D.~J.}\ \bibnamefont
  {Thouless}},\ }\href {\doibase 10.1088/0022-3719/6/7/010} {\bibfield
  {journal} {\bibinfo  {journal} {Journal of Physics C: Solid State Physics}\
  }\textbf {\bibinfo {volume} {6}},\ \bibinfo {pages} {1181} (\bibinfo {year}
  {1973})}\BibitemShut {NoStop}%
\bibitem [{\citenamefont {Agnolet}\ \emph {et~al.}(1989)\citenamefont
  {Agnolet}, \citenamefont {McQueeney},\ and\ \citenamefont
  {Reppy}}]{Agnolet1989}%
  \BibitemOpen
  \bibfield  {author} {\bibinfo {author} {\bibfnamefont {G.}~\bibnamefont
  {Agnolet}}, \bibinfo {author} {\bibfnamefont {D.~F.}\ \bibnamefont
  {McQueeney}}, \ and\ \bibinfo {author} {\bibfnamefont {J.~D.}\ \bibnamefont
  {Reppy}},\ }\href {\doibase 10.1103/PhysRevB.39.8934} {\bibfield  {journal}
  {\bibinfo  {journal} {Physical Review B}\ }\textbf {\bibinfo {volume} {39}},\
  \bibinfo {pages} {8934} (\bibinfo {year} {1989})}\BibitemShut {NoStop}%
\bibitem [{\citenamefont {Ceperley}\ and\ \citenamefont
  {Pollock}(1989)}]{Ceperley1989}%
  \BibitemOpen
  \bibfield  {author} {\bibinfo {author} {\bibfnamefont {D.~M.}\ \bibnamefont
  {Ceperley}}\ and\ \bibinfo {author} {\bibfnamefont {E.~L.}\ \bibnamefont
  {Pollock}},\ }\href {\doibase 10.1103/PhysRevB.39.2084} {\bibfield  {journal}
  {\bibinfo  {journal} {Physical Review B}\ }\textbf {\bibinfo {volume} {39}},\
  \bibinfo {pages} {2084} (\bibinfo {year} {1989})}\BibitemShut {NoStop}%
\bibitem [{\citenamefont {Gordillo}\ and\ \citenamefont
  {Ceperley}(1998)}]{Gordillo1998}%
  \BibitemOpen
  \bibfield  {author} {\bibinfo {author} {\bibfnamefont {M.}~\bibnamefont
  {Gordillo}}\ and\ \bibinfo {author} {\bibfnamefont {D.~M.}\ \bibnamefont
  {Ceperley}},\ }\href {\doibase 10.1103/PhysRevB.58.6447} {\bibfield
  {journal} {\bibinfo  {journal} {Physical Review B}\ }\textbf {\bibinfo
  {volume} {58}},\ \bibinfo {pages} {6447} (\bibinfo {year}
  {1998})}\BibitemShut {NoStop}%
\bibitem [{\citenamefont {Minnhagen}(1987)}]{Minnhagen1987}%
  \BibitemOpen
  \bibfield  {author} {\bibinfo {author} {\bibfnamefont {P.}~\bibnamefont
  {Minnhagen}},\ }\href {\doibase 10.1103/RevModPhys.59.1001} {\bibfield
  {journal} {\bibinfo  {journal} {Reviews of Modern Physics}\ }\textbf
  {\bibinfo {volume} {59}},\ \bibinfo {pages} {1001} (\bibinfo {year}
  {1987})}\BibitemShut {NoStop}%
\bibitem [{\citenamefont {Desbuquois}\ \emph {et~al.}(2012)\citenamefont
  {Desbuquois}, \citenamefont {Chomaz}, \citenamefont {Yefsah}, \citenamefont
  {L{\'{e}}onard}, \citenamefont {Beugnon}, \citenamefont {Weitenberg},\ and\
  \citenamefont {Dalibard}}]{Desbuquois2012}%
  \BibitemOpen
  \bibfield  {author} {\bibinfo {author} {\bibfnamefont {R.}~\bibnamefont
  {Desbuquois}}, \bibinfo {author} {\bibfnamefont {L.}~\bibnamefont {Chomaz}},
  \bibinfo {author} {\bibfnamefont {T.}~\bibnamefont {Yefsah}}, \bibinfo
  {author} {\bibfnamefont {J.}~\bibnamefont {L{\'{e}}onard}}, \bibinfo {author}
  {\bibfnamefont {J.}~\bibnamefont {Beugnon}}, \bibinfo {author} {\bibfnamefont
  {C.}~\bibnamefont {Weitenberg}}, \ and\ \bibinfo {author} {\bibfnamefont
  {J.}~\bibnamefont {Dalibard}},\ }\href {\doibase 10.1038/nphys2378}
  {\bibfield  {journal} {\bibinfo  {journal} {Nature Physics}\ }\textbf
  {\bibinfo {volume} {8}},\ \bibinfo {pages} {645} (\bibinfo {year}
  {2012})}\BibitemShut {NoStop}%
\bibitem [{\citenamefont {Ota}\ \emph {et~al.}(2018)\citenamefont {Ota},
  \citenamefont {Larcher}, \citenamefont {Dalfovo}, \citenamefont {Pitaevskii},
  \citenamefont {Proukakis},\ and\ \citenamefont {Stringari}}]{Ota2018}%
  \BibitemOpen
  \bibfield  {author} {\bibinfo {author} {\bibfnamefont {M.}~\bibnamefont
  {Ota}}, \bibinfo {author} {\bibfnamefont {F.}~\bibnamefont {Larcher}},
  \bibinfo {author} {\bibfnamefont {F.}~\bibnamefont {Dalfovo}}, \bibinfo
  {author} {\bibfnamefont {L.}~\bibnamefont {Pitaevskii}}, \bibinfo {author}
  {\bibfnamefont {N.~P.}\ \bibnamefont {Proukakis}}, \ and\ \bibinfo {author}
  {\bibfnamefont {S.}~\bibnamefont {Stringari}},\ }\href {\doibase
  10.1103/PhysRevLett.121.145302} {\bibfield  {journal} {\bibinfo  {journal}
  {Physical Review Letters}\ }\textbf {\bibinfo {volume} {121}},\ \bibinfo
  {pages} {145302} (\bibinfo {year} {2018})}\BibitemShut {NoStop}%
\bibitem [{\citenamefont {Murthy}\ \emph {et~al.}(2015)\citenamefont {Murthy},
  \citenamefont {Boettcher}, \citenamefont {Bayha}, \citenamefont {Holzmann},
  \citenamefont {Kedar}, \citenamefont {Neidig}, \citenamefont {Ries},
  \citenamefont {Wenz}, \citenamefont {Z{\"{u}}rn},\ and\ \citenamefont
  {Jochim}}]{Murthy2015}%
  \BibitemOpen
  \bibfield  {author} {\bibinfo {author} {\bibfnamefont {P.~A.}\ \bibnamefont
  {Murthy}}, \bibinfo {author} {\bibfnamefont {I.}~\bibnamefont {Boettcher}},
  \bibinfo {author} {\bibfnamefont {L.}~\bibnamefont {Bayha}}, \bibinfo
  {author} {\bibfnamefont {M.}~\bibnamefont {Holzmann}}, \bibinfo {author}
  {\bibfnamefont {D.}~\bibnamefont {Kedar}}, \bibinfo {author} {\bibfnamefont
  {M.}~\bibnamefont {Neidig}}, \bibinfo {author} {\bibfnamefont {M.~G.}\
  \bibnamefont {Ries}}, \bibinfo {author} {\bibfnamefont {A.~N.}\ \bibnamefont
  {Wenz}}, \bibinfo {author} {\bibfnamefont {G.}~\bibnamefont {Z{\"{u}}rn}}, \
  and\ \bibinfo {author} {\bibfnamefont {S.}~\bibnamefont {Jochim}},\ }\href
  {\doibase 10.1103/PhysRevLett.115.010401} {\bibfield  {journal} {\bibinfo
  {journal} {Physical Review Letters}\ }\textbf {\bibinfo {volume} {115}},\
  \bibinfo {pages} {010401} (\bibinfo {year} {2015})}\BibitemShut {NoStop}%
\bibitem [{\citenamefont {Carleo}\ \emph {et~al.}(2013)\citenamefont {Carleo},
  \citenamefont {Bo{\'{e}}ris}, \citenamefont {Holzmann},\ and\ \citenamefont
  {Sanchez-Palencia}}]{Carleo2013}%
  \BibitemOpen
  \bibfield  {author} {\bibinfo {author} {\bibfnamefont {G.}~\bibnamefont
  {Carleo}}, \bibinfo {author} {\bibfnamefont {G.}~\bibnamefont
  {Bo{\'{e}}ris}}, \bibinfo {author} {\bibfnamefont {M.}~\bibnamefont
  {Holzmann}}, \ and\ \bibinfo {author} {\bibfnamefont {L.}~\bibnamefont
  {Sanchez-Palencia}},\ }\href {\doibase 10.1103/PhysRevLett.111.050406}
  {\bibfield  {journal} {\bibinfo  {journal} {Physical Review Letters}\
  }\textbf {\bibinfo {volume} {111}},\ \bibinfo {pages} {050406} (\bibinfo
  {year} {2013})}\BibitemShut {NoStop}%
\bibitem [{\citenamefont {Maccari}\ \emph {et~al.}(2018)\citenamefont
  {Maccari}, \citenamefont {Benfatto}, \citenamefont {Castellani},
  \citenamefont {Maccari}, \citenamefont {Benfatto},\ and\ \citenamefont
  {Castellani}}]{Maccari2018}%
  \BibitemOpen
  \bibfield  {author} {\bibinfo {author} {\bibfnamefont {I.}~\bibnamefont
  {Maccari}}, \bibinfo {author} {\bibfnamefont {L.}~\bibnamefont {Benfatto}},
  \bibinfo {author} {\bibfnamefont {C.}~\bibnamefont {Castellani}}, \bibinfo
  {author} {\bibfnamefont {I.}~\bibnamefont {Maccari}}, \bibinfo {author}
  {\bibfnamefont {L.}~\bibnamefont {Benfatto}}, \ and\ \bibinfo {author}
  {\bibfnamefont {C.}~\bibnamefont {Castellani}},\ }\href {\doibase
  10.3390/condmat3010008} {\bibfield  {journal} {\bibinfo  {journal} {Condensed
  Matter}\ }\textbf {\bibinfo {volume} {3}},\ \bibinfo {pages} {8} (\bibinfo
  {year} {2018})}\BibitemShut {NoStop}%
\bibitem [{\citenamefont {Filinov}\ \emph {et~al.}(2010)\citenamefont
  {Filinov}, \citenamefont {Prokof'Ev},\ and\ \citenamefont
  {Bonitz}}]{Filinov2010}%
  \BibitemOpen
  \bibfield  {author} {\bibinfo {author} {\bibfnamefont {A.}~\bibnamefont
  {Filinov}}, \bibinfo {author} {\bibfnamefont {N.~V.}\ \bibnamefont
  {Prokof'Ev}}, \ and\ \bibinfo {author} {\bibfnamefont {M.}~\bibnamefont
  {Bonitz}},\ }\href {\doibase 10.1103/PhysRevLett.105.070401} {\bibfield
  {journal} {\bibinfo  {journal} {Physical Review Letters}\ }\textbf {\bibinfo
  {volume} {105}},\ \bibinfo {pages} {070401} (\bibinfo {year}
  {2010})}\BibitemShut {NoStop}%
\bibitem [{\citenamefont {Astrakharchik}\ \emph {et~al.}(2007)\citenamefont
  {Astrakharchik}, \citenamefont {Boronat}, \citenamefont {Kurbakov},\ and\
  \citenamefont {Lozovik}}]{Astrakharchik2007}%
  \BibitemOpen
  \bibfield  {author} {\bibinfo {author} {\bibfnamefont {G.~E.}\ \bibnamefont
  {Astrakharchik}}, \bibinfo {author} {\bibfnamefont {J.}~\bibnamefont
  {Boronat}}, \bibinfo {author} {\bibfnamefont {I.~L.}\ \bibnamefont
  {Kurbakov}}, \ and\ \bibinfo {author} {\bibfnamefont {Y.~E.}\ \bibnamefont
  {Lozovik}},\ }\href {\doibase 10.1103/PhysRevLett.98.060405} {\bibfield
  {journal} {\bibinfo  {journal} {Physical Review Letters}\ }\textbf {\bibinfo
  {volume} {98}},\ \bibinfo {pages} {060405} (\bibinfo {year}
  {2007})}\BibitemShut {NoStop}%
\bibitem [{\citenamefont {Ceperley}(1995)}]{Ceperley1995}%
  \BibitemOpen
  \bibfield  {author} {\bibinfo {author} {\bibfnamefont {D.~M.}\ \bibnamefont
  {Ceperley}},\ }\href {\doibase 10.1103/RevModPhys.67.279} {\bibfield
  {journal} {\bibinfo  {journal} {Reviews of Modern Physics}\ }\textbf
  {\bibinfo {volume} {67}},\ \bibinfo {pages} {279} (\bibinfo {year}
  {1995})}\BibitemShut {NoStop}%
\bibitem [{\citenamefont {Takahashi}\ and\ \citenamefont
  {Imada}(1984{\natexlab{a}})}]{Takahashi1984}%
  \BibitemOpen
  \bibfield  {author} {\bibinfo {author} {\bibfnamefont {M.}~\bibnamefont
  {Takahashi}}\ and\ \bibinfo {author} {\bibfnamefont {M.}~\bibnamefont
  {Imada}},\ }\href {\doibase 10.1143/JPSJ.53.963} {\bibfield  {journal}
  {\bibinfo  {journal} {Journal of the Physical Society of Japan}\ }\textbf
  {\bibinfo {volume} {53}},\ \bibinfo {pages} {963} (\bibinfo {year}
  {1984}{\natexlab{a}})}\BibitemShut {NoStop}%
\bibitem [{\citenamefont {Takahashi}\ and\ \citenamefont
  {Imada}(1984{\natexlab{b}})}]{Takahashi1984a}%
  \BibitemOpen
  \bibfield  {author} {\bibinfo {author} {\bibfnamefont {M.}~\bibnamefont
  {Takahashi}}\ and\ \bibinfo {author} {\bibfnamefont {M.}~\bibnamefont
  {Imada}},\ }\href {\doibase 10.1143/JPSJ.53.3765} {\bibfield  {journal}
  {\bibinfo  {journal} {Journal of the Physical Society of Japan}\ }\textbf
  {\bibinfo {volume} {53}},\ \bibinfo {pages} {3765} (\bibinfo {year}
  {1984}{\natexlab{b}})}\BibitemShut {NoStop}%
\bibitem [{\citenamefont {Chin}(2004)}]{Chin2004}%
  \BibitemOpen
  \bibfield  {author} {\bibinfo {author} {\bibfnamefont {S.~A.}\ \bibnamefont
  {Chin}},\ }\href {\doibase 10.1103/PhysRevE.69.046118} {\bibfield  {journal}
  {\bibinfo  {journal} {Physical Review E}\ }\textbf {\bibinfo {volume} {69}},\
  \bibinfo {pages} {046118} (\bibinfo {year} {2004})}\BibitemShut {NoStop}%
\bibitem [{\citenamefont {Chin}\ and\ \citenamefont {Chen}(2002)}]{Chin2002}%
  \BibitemOpen
  \bibfield  {author} {\bibinfo {author} {\bibfnamefont {S.~A.}\ \bibnamefont
  {Chin}}\ and\ \bibinfo {author} {\bibfnamefont {C.~R.}\ \bibnamefont
  {Chen}},\ }\href {\doibase 10.1063/1.1485725} {\bibfield  {journal} {\bibinfo
   {journal} {The Journal of Chemical Physics}\ }\textbf {\bibinfo {volume}
  {117}},\ \bibinfo {pages} {1409} (\bibinfo {year} {2002})}\BibitemShut
  {NoStop}%
\bibitem [{\citenamefont {Sakkos}\ \emph {et~al.}(2009)\citenamefont {Sakkos},
  \citenamefont {Casulleras},\ and\ \citenamefont {Boronat}}]{Sakkos2009}%
  \BibitemOpen
  \bibfield  {author} {\bibinfo {author} {\bibfnamefont {K.}~\bibnamefont
  {Sakkos}}, \bibinfo {author} {\bibfnamefont {J.}~\bibnamefont {Casulleras}},
  \ and\ \bibinfo {author} {\bibfnamefont {J.}~\bibnamefont {Boronat}},\ }\href
  {\doibase 10.1063/1.3143522} {\bibfield  {journal} {\bibinfo  {journal}
  {Journal of Chemical Physics}\ }\textbf {\bibinfo {volume} {130}} (\bibinfo
  {year} {2009}),\ 10.1063/1.3143522},\ \Eprint
  {http://arxiv.org/abs/0903.2763} {0903.2763} \BibitemShut {NoStop}%
\bibitem [{\citenamefont {Boninsegni}\ \emph {et~al.}(2006)\citenamefont
  {Boninsegni}, \citenamefont {Prokof'ev},\ and\ \citenamefont
  {Svistunov}}]{Boninsegni2006}%
  \BibitemOpen
  \bibfield  {author} {\bibinfo {author} {\bibfnamefont {M.}~\bibnamefont
  {Boninsegni}}, \bibinfo {author} {\bibfnamefont {N.~V.}\ \bibnamefont
  {Prokof'ev}}, \ and\ \bibinfo {author} {\bibfnamefont {B.~V.}\ \bibnamefont
  {Svistunov}},\ }\href {\doibase 10.1103/PhysRevE.74.036701} {\bibfield
  {journal} {\bibinfo  {journal} {Physical Review E}\ }\textbf {\bibinfo
  {volume} {74}},\ \bibinfo {pages} {036701} (\bibinfo {year}
  {2006})}\BibitemShut {NoStop}%
\bibitem [{\citenamefont {Nelson}\ and\ \citenamefont
  {Kosterlitz}(1977)}]{Nelson1977}%
  \BibitemOpen
  \bibfield  {author} {\bibinfo {author} {\bibfnamefont {D.~R.}\ \bibnamefont
  {Nelson}}\ and\ \bibinfo {author} {\bibfnamefont {J.~M.}\ \bibnamefont
  {Kosterlitz}},\ }\href {\doibase 10.1103/PhysRevLett.39.1201} {\bibfield
  {journal} {\bibinfo  {journal} {Physical Review Letters}\ }\textbf {\bibinfo
  {volume} {39}},\ \bibinfo {pages} {1201} (\bibinfo {year}
  {1977})}\BibitemShut {NoStop}%
\bibitem [{\citenamefont {Kosterlitz}(1974)}]{Kosterlitz1974}%
  \BibitemOpen
  \bibfield  {author} {\bibinfo {author} {\bibfnamefont {J.~M.}\ \bibnamefont
  {Kosterlitz}},\ }\href {\doibase 10.1088/0022-3719/7/6/005} {\bibfield
  {journal} {\bibinfo  {journal} {J. Phys. C: Solid State Phys.}\ }\textbf
  {\bibinfo {volume} {7}},\ \bibinfo {pages} {1046} (\bibinfo {year}
  {1974})}\BibitemShut {NoStop}%
\bibitem [{\citenamefont {Pollock}\ and\ \citenamefont
  {Ceperley}(1987)}]{Pollock1987}%
  \BibitemOpen
  \bibfield  {author} {\bibinfo {author} {\bibfnamefont {E.~L.}\ \bibnamefont
  {Pollock}}\ and\ \bibinfo {author} {\bibfnamefont {D.~M.}\ \bibnamefont
  {Ceperley}},\ }\href {\doibase 10.1103/PhysRevB.36.8343} {\bibfield
  {journal} {\bibinfo  {journal} {Physical Review B}\ }\textbf {\bibinfo
  {volume} {36}},\ \bibinfo {pages} {8343} (\bibinfo {year}
  {1987})}\BibitemShut {NoStop}%
\bibitem [{\citenamefont {Macia}\ \emph {et~al.}(2014)\citenamefont {Macia},
  \citenamefont {Boronat},\ and\ \citenamefont {Mazzanti}}]{Macia2014}%
  \BibitemOpen
  \bibfield  {author} {\bibinfo {author} {\bibfnamefont {A.}~\bibnamefont
  {Macia}}, \bibinfo {author} {\bibfnamefont {J.}~\bibnamefont {Boronat}}, \
  and\ \bibinfo {author} {\bibfnamefont {F.}~\bibnamefont {Mazzanti}},\ }\href
  {\doibase 10.1103/PhysRevA.90.061601} {\bibfield  {journal} {\bibinfo
  {journal} {Physical Review A - Atomic, Molecular, and Optical Physics}\
  }\textbf {\bibinfo {volume} {90}},\ \bibinfo {pages} {1} (\bibinfo {year}
  {2014})}\BibitemShut {NoStop}%
\bibitem [{\citenamefont {Mazzanti}\ \emph {et~al.}(2009)\citenamefont
  {Mazzanti}, \citenamefont {Zillich}, \citenamefont {Astrakharchik},\ and\
  \citenamefont {Boronat}}]{Macia2009}%
  \BibitemOpen
  \bibfield  {author} {\bibinfo {author} {\bibfnamefont {F.}~\bibnamefont
  {Mazzanti}}, \bibinfo {author} {\bibfnamefont {R.~E.}\ \bibnamefont
  {Zillich}}, \bibinfo {author} {\bibfnamefont {G.~E.}\ \bibnamefont
  {Astrakharchik}}, \ and\ \bibinfo {author} {\bibfnamefont {J.}~\bibnamefont
  {Boronat}},\ }\href {\doibase 10.1103/PhysRevLett.102.110405} {\bibfield
  {journal} {\bibinfo  {journal} {Physical Review Letters}\ }\textbf {\bibinfo
  {volume} {102}},\ \bibinfo {pages} {110405} (\bibinfo {year}
  {2009})}\BibitemShut {NoStop}%
\bibitem [{\citenamefont {Macia}\ \emph {et~al.}(2011)\citenamefont {Macia},
  \citenamefont {Mazzanti}, \citenamefont {Boronat},\ and\ \citenamefont
  {Zillich}}]{Macia2011}%
  \BibitemOpen
  \bibfield  {author} {\bibinfo {author} {\bibfnamefont {A.}~\bibnamefont
  {Macia}}, \bibinfo {author} {\bibfnamefont {F.}~\bibnamefont {Mazzanti}},
  \bibinfo {author} {\bibfnamefont {J.}~\bibnamefont {Boronat}}, \ and\
  \bibinfo {author} {\bibfnamefont {R.~E.}\ \bibnamefont {Zillich}},\ }\href
  {\doibase 10.1103/PhysRevA.84.033625} {\bibfield  {journal} {\bibinfo
  {journal} {Phys. Rev. A}\ }\textbf {\bibinfo {volume} {84}},\ \bibinfo
  {pages} {033625} (\bibinfo {year} {2011})}\BibitemShut {NoStop}%
\bibitem [{\citenamefont {{Vranje{\v{s}} Marki{\'{c}}}}\ \emph
  {et~al.}(2018)\citenamefont {{Vranje{\v{s}} Marki{\'{c}}}}, \citenamefont
  {Vrcan}, \citenamefont {Zuhrianda},\ and\ \citenamefont
  {Glyde}}]{VranjesMarkic2018}%
  \BibitemOpen
  \bibfield  {author} {\bibinfo {author} {\bibfnamefont {L.}~\bibnamefont
  {{Vranje{\v{s}} Marki{\'{c}}}}}, \bibinfo {author} {\bibfnamefont
  {H.}~\bibnamefont {Vrcan}}, \bibinfo {author} {\bibfnamefont
  {Z.}~\bibnamefont {Zuhrianda}}, \ and\ \bibinfo {author} {\bibfnamefont
  {H.~R.}\ \bibnamefont {Glyde}},\ }\href {\doibase 10.1103/PhysRevB.97.014513}
  {\bibfield  {journal} {\bibinfo  {journal} {Physical Review B}\ }\textbf
  {\bibinfo {volume} {97}},\ \bibinfo {pages} {014513} (\bibinfo {year}
  {2018})}\BibitemShut {NoStop}%
\bibitem [{\citenamefont {Macia}\ \emph {et~al.}(2012)\citenamefont {Macia},
  \citenamefont {Hufnagl}, \citenamefont {Mazzanti}, \citenamefont {Boronat},\
  and\ \citenamefont {Zillich}}]{Macia2012}%
  \BibitemOpen
  \bibfield  {author} {\bibinfo {author} {\bibfnamefont {A.}~\bibnamefont
  {Macia}}, \bibinfo {author} {\bibfnamefont {D.}~\bibnamefont {Hufnagl}},
  \bibinfo {author} {\bibfnamefont {F.}~\bibnamefont {Mazzanti}}, \bibinfo
  {author} {\bibfnamefont {J.}~\bibnamefont {Boronat}}, \ and\ \bibinfo
  {author} {\bibfnamefont {R.~E.}\ \bibnamefont {Zillich}},\ }\href {\doibase
  10.1103/PhysRevLett.109.235307} {\bibfield  {journal} {\bibinfo  {journal}
  {Physical Review Letters}\ }\textbf {\bibinfo {volume} {109}},\ \bibinfo
  {pages} {235307} (\bibinfo {year} {2012})}\BibitemShut {NoStop}%
\bibitem [{\citenamefont {Wu}\ \emph {et~al.}(2016)\citenamefont {Wu},
  \citenamefont {Block},\ and\ \citenamefont {Bruun}}]{wu2016liquid}%
  \BibitemOpen
  \bibfield  {author} {\bibinfo {author} {\bibfnamefont {Z.}~\bibnamefont
  {Wu}}, \bibinfo {author} {\bibfnamefont {J.~K.}\ \bibnamefont {Block}}, \
  and\ \bibinfo {author} {\bibfnamefont {G.~M.}\ \bibnamefont {Bruun}},\
  }\href@noop {} {\bibfield  {journal} {\bibinfo  {journal} {Scientific
  Reports}\ }\textbf {\bibinfo {volume} {6}},\ \bibinfo {pages} {19038}
  (\bibinfo {year} {2016})}\BibitemShut {NoStop}%
\bibitem [{\citenamefont {Mendoza-Coto}\ \emph {et~al.}(2017)\citenamefont
  {Mendoza-Coto}, \citenamefont {Barci},\ and\ \citenamefont
  {Stariolo}}]{mendoza2017quantum}%
  \BibitemOpen
  \bibfield  {author} {\bibinfo {author} {\bibfnamefont {A.}~\bibnamefont
  {Mendoza-Coto}}, \bibinfo {author} {\bibfnamefont {D.~G.}\ \bibnamefont
  {Barci}}, \ and\ \bibinfo {author} {\bibfnamefont {D.~A.}\ \bibnamefont
  {Stariolo}},\ }\href@noop {} {\bibfield  {journal} {\bibinfo  {journal}
  {Physical Review B}\ }\textbf {\bibinfo {volume} {95}},\ \bibinfo {pages}
  {144209} (\bibinfo {year} {2017})}\BibitemShut {NoStop}%
\bibitem [{\citenamefont {Masella}\ \emph {et~al.}(2019)\citenamefont
  {Masella}, \citenamefont {Angelone}, \citenamefont {Mezzacapo}, \citenamefont
  {Pupillo},\ and\ \citenamefont {Prokof'ev}}]{Masella2019}%
  \BibitemOpen
  \bibfield  {author} {\bibinfo {author} {\bibfnamefont {G.}~\bibnamefont
  {Masella}}, \bibinfo {author} {\bibfnamefont {A.}~\bibnamefont {Angelone}},
  \bibinfo {author} {\bibfnamefont {F.}~\bibnamefont {Mezzacapo}}, \bibinfo
  {author} {\bibfnamefont {G.}~\bibnamefont {Pupillo}}, \ and\ \bibinfo
  {author} {\bibfnamefont {N.~V.}\ \bibnamefont {Prokof'ev}},\ }\href {\doibase
  10.1103/PhysRevLett.123.045301} {\bibfield  {journal} {\bibinfo  {journal}
  {Phys. Rev. Lett.}\ }\textbf {\bibinfo {volume} {123}},\ \bibinfo {pages}
  {045301} (\bibinfo {year} {2019})}\BibitemShut {NoStop}%
\end{thebibliography}%

\end{document}